%% file: main.tex
\documentclass[10pt,twocolumn,letterpaper]{article}

 \usepackage[pagenumbers]{cvpr} 


\definecolor{cvprblue}{rgb}{0.21,0.49,0.74}
\usepackage[pagebackref,breaklinks,colorlinks,allcolors=cvprblue]{hyperref}
\makeatletter
\renewcommand\paragraph{\@startsection{paragraph}{4}{\z@}%
                                    {0.4\baselineskip} 
                                    {-1em} 
                                    {\normalfont\normalsize\bfseries}}
\makeatother

\def\paperID{2987} 
\def\confName{CVPR}
\def\confYear{2025}

\newcommand{\name}{\textit{Lux Post Facto}\xspace}

\title{Lux Post Facto: Learning Portrait Performance Relighting with \\ Conditional Video Diffusion and a Hybrid Dataset}

\author{Yiqun~Mei$^{1,2}$\quad Mingming He$^{1}$\quad Li Ma$^{1}$ \quad Julien Philip$^{1}$\quad Wenqi Xian$^{1}$\quad David M George$^{1}$\\ Xueming Yu$^{1}$\quad Gabriel Dedic$^{1}$\quad Ahmet Levent Taşel$^{1}$\quad Ning Yu$^{1}$\quad Vishal M.~Patel$^{2}$\quad Paul Debevec$^{1}$ \\
  {\small$^{1}$ Netflix Eyeline Studios\quad \quad $^{2}$Johns Hopkins University}
}

\begin{document}
\twocolumn[{%
\renewcommand\twocolumn[1][]{#1}%
\maketitle
\begin{center}
    \centering
    \captionsetup{type=figure}
    \vspace{-1.8em}
    
    \makebox[0.162\linewidth]{\centering \small Input}
    \hfill
    \makebox[0.162\linewidth]{\centering \small ``Vestibule''}
    \hfill
    \makebox[0.162\linewidth]{\centering \small ``Cape Hill''}
    \hfill
    \makebox[0.162\linewidth]{\centering \small ``Park Parking''}
    \hfill
    \makebox[0.162\linewidth]{\centering \small Point Lights  }
    \hfill
    \makebox[0.162\linewidth]{\centering \small  Rim Light}
    
    \includegraphics[width=\linewidth]{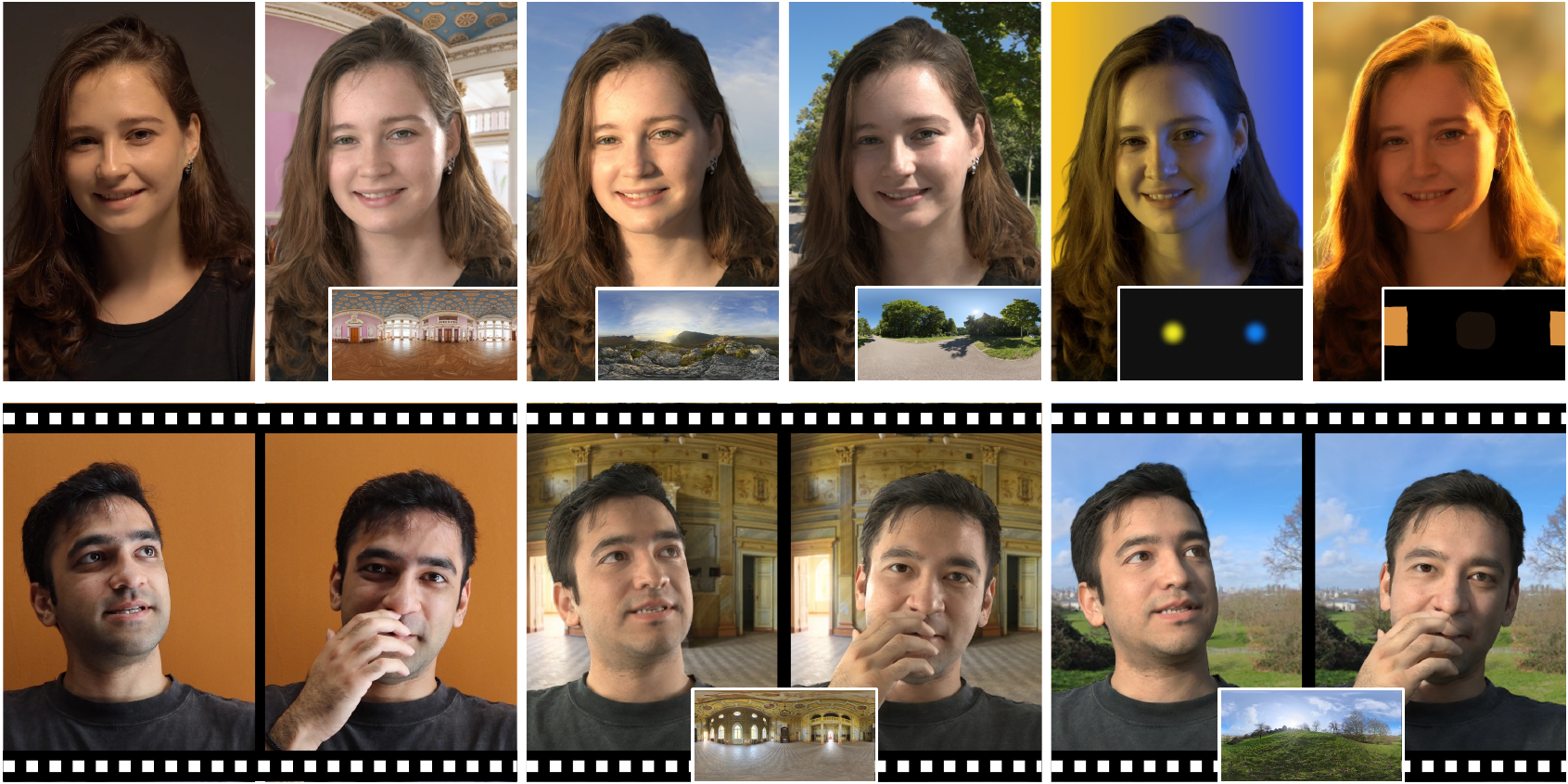}

    \vspace{-1mm}
    \makebox[0.33\linewidth]{\centering \small Input Video}
    \hfill
    \makebox[0.33\linewidth]{\centering \small ``Old Hall''}
    \hfill
    \makebox[0.33\linewidth]{\centering \small ``Greenwich Park''}
    
\vskip-5pt \captionof{figure}{\textbf{\name} offers portrait relighting as a simple post-production process. Users can edit the lighting of portrait images (first row) and videos (second row) with high fidelity using any HDR map. Our method is temporally stable and highly photorealistic.}
    \label{fig:teaser}

\end{center}
}]
\input{sec/0_abstract}    
\input{sec/1_intro}
\input{sec/2_related}

\input{sec/3_method}

\input{sec/4_data}

\input{sec/5_experiment}

\input{sec/6_conclusion}

{
    \small
    \bibliographystyle{ieeenat_fullname}
    \bibliography{main}
}
\appendix
\clearpage
\input{supp}

\end{document}

%% file: sec/0_abstract.tex
\begin{abstract}

Video portrait relighting remains challenging because the results need to be both photorealistic and temporally stable.
This typically requires a strong model design that can capture complex facial reflections as well as intensive training on a high-quality paired video dataset, such as \textit{dynamic} one-light-at-a-time (OLAT).
In this work, we introduce \textbf{\name}, a novel portrait video relighting method that produces both photorealistic and temporally consistent lighting effects.
From the model side, we design a new conditional video diffusion model built upon state-of-the-art pre-trained video diffusion model, alongside a new lighting injection mechanism to enable precise control.
This way we leverage strong spatial and temporal generative capability to generate plausible solutions to the ill-posed relighting problem.
Our technique uses a hybrid dataset consisting of static expression OLAT data and in-the-wild portrait performance videos to jointly learn relighting and temporal modeling.
This avoids the need to acquire paired video data in different lighting conditions.
Our extensive experiments show that our model produces state-of-the-art results both in terms of photorealism and temporal consistency. Video results can be found on our \href{https://www.eyelinestudios.com/research/luxpostfacto.html}{project page}.

\end{abstract}

%% file: sec/1_intro.tex
\section{Introduction}
\label{sec:intro}

Lighting is a crucial element in visual storytelling, shaping the scene, adding depth, and creating emotional impact.
Because of this, lighting receives considerable attention in the filming process, marked by professional expertise, expensive equipment, and coordinated teamwork -- resources beyond the reach of novice users.
In this work, we endeavor to design a video relighting method to empower everyday content creators to create creative and compelling lighting in portrait videos as a simple post-production process. 

Video relighting requires an accurate simulation of light transport through complex materials across both time and space.
Image-based relighting \cite{reflectance_field_paul} allows a portrait to be relit when the subject has been recorded from a dense array of lighting directions, such as a one-light-at-a-time (OLAT) reflectance field. Using a high-speed camera and a synchronized LED stage, \citet{performance_relighting_paul} record OLATs at movie frame rates, allowing for cinematic relighting of a facial performance. However, the technique relies on complex equipment and cannot be generalized to new subjects.

More recent works have attempted to transfer facial reflectance information from one or more subjects to another, for example by tracking an OLAT quotient image \cite{quotient_image_paul} onto a new flat-lit performance.
This has been generalized to a deep learning context, where OLAT data from a set of training subjects is used to relight a novel subject~\cite{single_portrait_relighting,total_relighting,nvpr,switchlight,uravatar}.
However, these models can have issues with temporal consistency, a lack of photorealism, or require difficult-to-obtain OLAT video data, or some combination of the three.

In this work, we propose \textbf{\name}, a new video relighting model for relighting arbitrary portrait videos realistically and with temporal consistency. We formulate video relighting as a conditional generation process leveraging a state-of-the-art video diffusion model~\cite{stable_video_diffusion}, fine-tuned on a hybrid dataset with new lighting injection
and training strategies.

While video diffusion models~\cite{stable_video_diffusion, cogvideox} have the power to generate temporally consistent videos with high-quality lighting from text guidance, they lack the ability to achieve fine-grain lighting control.
While we would like to specify the lighting as a high dynamic range (HDR) lighting environment, conventional image conditioning methods~\cite{ye2023ip,zhao2024uni} like CLIP~\cite{CLIP}, struggle to represent the HDR pixel values and spatial details in HDR maps.
And methods based on pixel-aligned control maps \cite{controlnet,lightit} require explicit proxy geometry, prone to estimation errors.

We solve this problem by encoding the original HDR map as a set of ``lighting embeddings'', where each embedding represents a single directional light source.
Collectively, the embeddings produce a complete lighting environment representation.
We pass this representation to the diffusion model through cross-attention to achieve precise lighting control.

Our model is practical to train since it requires only a limited number of \textit{static} OLAT image datasets and a larger set of in-the-wild portrait videos.
The static OLATs provide relighting supervision for individual frames, and the in-the-wild videos train the model to produce temporally stable performances under a wild variety of unknown lighting conditions.
Our novel training strategy learns from this combination of datasets to achieve temporally consistent portrait video relighting. We show that our method outperforms current single-image and video relighting models.

To summarize, our technical contributions include:
\begin{enumerate}     
     \item A novel video diffusion model for portrait relighting, capable of generating high-fidelity lighting effects on arbitrary portrait videos with state-of-the-art performance.     
     \item A novel lighting control module that improves the encoding of lighting information and enables precise lighting control.     
     \item A new hybrid dataset and associated training approach which allows training our video relighting models using static OLAT images and in-the-wild video data in a unified framework. 
 \end{enumerate}

%% file: sec/2_related.tex
\section{Related Work}
\label{sec:related}

\begin{figure*}[t!]
    \centering
    \includegraphics[width=1.0\linewidth]{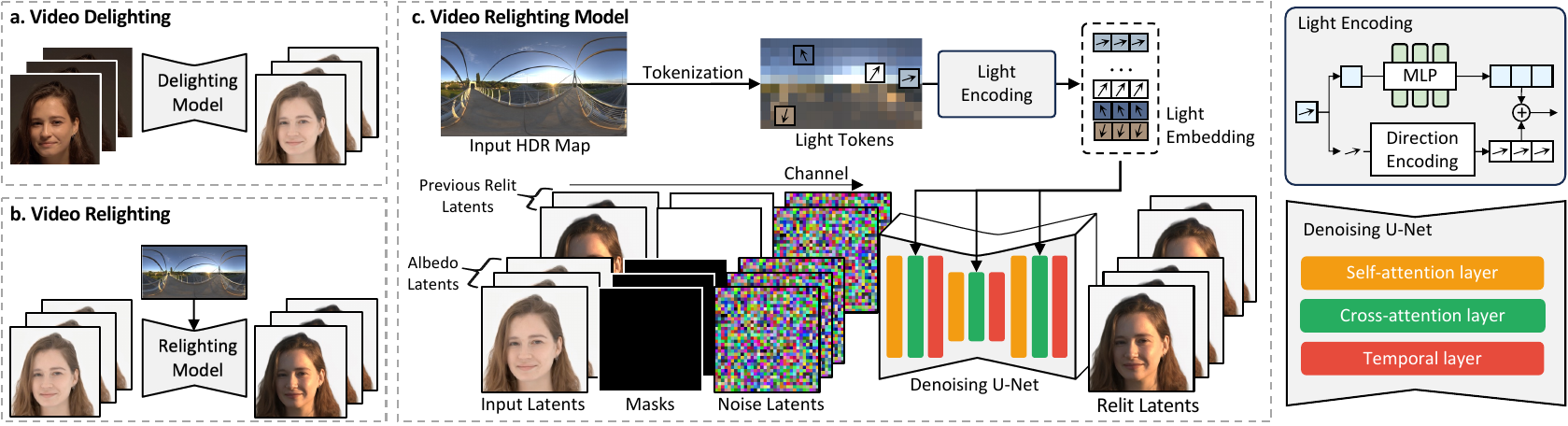}
    \vspace{-5mm}
    \caption{\textbf{Overview and model design of \name.} To relight an input video, a delighting model predicts an albedo video (a) which is then relit by a relighting model (b).
    Both models share the same architecture (c) based on stable video diffusion \cite{stable_video_diffusion} (SVD).
    We condition the SVD on the input video by concatenating input latents to the Gaussian noise.
    To support autoregressive prediction for long sequence, we replace the first $T$ frames with previous predictions, indicated with a binary mask concatenated to the input.
    The output lighting is controlled by an HDR map, converted to a light embedding fed to the U-Net through cross-attention layers.
    The VAE that encodes and decodes the latents is omitted for clarity.}
    \label{fig:pipeline}
    \vspace{-5mm}
\end{figure*}

Portrait relighting has been a key research area for years.
The pioneering light stage work~\cite{olat_paul} proposes to capture the reflectance field of human faces using OLAT, composing them to novel lighting using the linearity of light transport.
Ensuing work~\cite{performance_relighting_paul,reflectance_field_paul,the_relightables,deep_reflectance_field,deep_relightable_texture,deep_appearance_models} use time-multiplexed lighting to capture live-action actors at high frame rates.
Moving subjects are then relit using image-based or neural rendering~\cite{ibr_face,neural_rendering}.
Alternative multi-view setups (\eg,~\citet{sun_stage}) have been proposed to create relighting models.
However, all these methods require per-subject capture, limiting their generalization.

To bypass the need for light stage data, techniques like quotient images~\cite{quotient_image,quotient_image_paul}, intrinsic decomposition~\cite{shape_illu_reflect,face_relighting_consistent,illumination_invariant,intrinsic_face_decomp,join_julien,inverse_lighting}, or style transfer~\cite{shih2014,mass_transport_relight} have been proposed.
Recently, data-driven methods use models trained on multi-illumination datasets to generalize.
CNN-based methods have shown their effectiveness on image relighting, achieving novel lighting effects~\cite{indoor_relighting_julien,total_relighting,switchlight,lumos}, shadow manipulation~\cite{single_portrait_relighting,deep_single,shadow_manip,outdoor_relighting_julien,outcast,portrait_relight_channel,face_relighting_consistent,smallot, shadow1,shadow2,shadow3}, and interactive lighting editing~\cite{lightpainter}.
GAN-based methods enhance portrait relighting using ratio images~\cite{towards_face_relight}, or explicit 3D representations~\cite{volux_gan,holo_relight,facelit, shading_guided_gen}.
While effective for single images, these methods struggle to maintain temporal consistency when applied to videos.

Given the recent advances in image diffusion models~\cite{diff_beat_gan,latent_diff}, known for their quality and generality, some works apply them to portrait and scene relighting~\cite{difareli,lightit,dlightNet,diff_approach_julien,diffrelight, hou_shadow}, and related tasks like portrait harmonization~\cite{iclight,relight_harmonization}.
The relighting models~\cite{lightit,dlightNet,diff_approach_julien,diffrelight} fine-tune pre-trained latent diffusion models~\cite{latent_diff} on a dataset with lighting annotations.
The harmonization methods~\cite{iclight,relight_harmonization} propose to adjust foreground lighting to match backgrounds.
They both achieve photorealistic results.
However, none of these image-based models are temporally stable.

\paragraph{Video-based Relighting.}
Video relighting focuses on consistent lighting for dynamic sequences.
Recent methods often trade spatial quality for temporal consistency.
For example, extensions of image relighting models to video, such as~\cite{switchlight,diffrelight}, typically apply temporal smoothing to the relit results, leading to blurry shading and averaged details. Some other methods assume simplified lighting conditions, such as diffuse~\cite{EdgeRelight360}, or white and spherical harmonic illumination~\cite{cai_etal}.
NVPR~\cite{nvpr}, a general face relighting model, integrates a temporal consistency loss to achieve stable portrait relighting under dynamic lighting, but loses some facial details.
Some 3D-based methods~\cite{vrmm,uravatar} leverage the advanced 3DGS representations~\cite{3DGS} for high-quality facial performance relighting.
They require per-subject multi-view capture for both training and inference.
These methods rely on scarce dynamic OLAT datasets captured with high-speed cameras.
We introduce a hybrid dataset to address the scarcity of high-quality paired video datasets.

\paragraph{Video Diffusion Models.}
Video diffusion models~\cite{align_latent,make_a_video,animate_diff,stable_video_diffusion} form the foundation of our video relighting model.
These models synthesize realistic and temporally coherent videos from text input.
To improve the generation quality, some works propose to use diffusion noise prior~\cite{warp_noise,preserve_correlation} or cascading models~\cite{imagen_video,lavie}.
To make the model controllable, recent works condition it on control signals~\cite{control_video} such as human poses~\cite{magic_pose,video_composer}, input images~\cite{marigold_video}, or depth maps~\cite{content_guided}.
To precisely control the lighting of our video diffusion models, we propose a new light conditioning mechanism. 

\paragraph{Preliminary:} Our video relighting model is built upon video diffusion models like Stable Video Diffusion (SVD)~\cite{stable_video_diffusion}. These models involve a forward pass that progressively injects Gaussian noise into video sequences and a reverse process with a denoising U-Net learned to predict the noise to reconstruct the original videos. The network is trained by minimizing the mean squared error between noise predictions and ground truth. To ensure temporal coherence, the models incorporate specialized layers like 3D convolutions and temporal attention, operating in a latent space via a variational autoencoder (VAE) for efficiency.

%% file: sec/3_method.tex
\section{Method}
\label{sec:method}

We present \name, a video relighting method that can relight in-the-wild portrait videos.
Following recent work on portrait relighting \cite{total_relighting,holo_relight}, our method is composed of two stages: a video delighting pass and a video relighting pass.
Both stages are shown in Fig.~\ref{fig:pipeline}.a and ~\ref{fig:pipeline}.b.
First, the delighting model takes a portrait video as input and predicts a shading-free albedo\footnote{Following~\cite{total_relighting,lightpainter, holo_relight}, we refer to flat-lit images as albedo approximation.} video.
Then, the relighting model takes the predicted albedo video as input and an HDR map as lighting condition and generates a relit video under the target lighting condition.
The two models are built upon the same video diffusion architecture but differ in input and conditioning.

We train both these models on a hybrid dataset stemming from static OLAT images with precise lighting information and in-the-wild portrait videos with diverse motions.
To best use this data, we train our model with the auxiliary task of reference-based appearance copy, improving temporal consistency.

In Sec.~\ref{ssec:model}, we first describe both the relighting and delighting model design and our long-sequence inference strategy.
Then in Sec.~\ref{ssec:model_training}, we detail our hybrid dataset and how we adapt them to the auxiliary task of appearance copy.

\subsection{Model Design}\label{ssec:model}

At the core of \name is the idea to leverage the strong generative priors from pretrained video diffusion models, adapting them for the relighting task.
We thus build our relighting method upon a state-of-the-art open-source video diffusion model, \ie, SVD~\cite{stable_video_diffusion}, modified to become a conditional generator.
To achieve our relighting goal, we propose a two-stage approach following existing works~\cite{total_relighting,holo_relight}, which first delights the input video to generate its delit albedo video and then relights it.
Since both modules share the same backbone, we first introduce the relighting model, then describe how the delighting model differs.
Note that both models are trained independently and do not share their weights.

\subsubsection{Relighting Model}

The model architecture of the relighting module is shown in Fig.~\ref{fig:pipeline}.c.
From a delit albedo video clip - predicted by the delighting module - and potentially previous relit frames, the relighting model predicts a relit video.
To adapt the text-guided SVD model to our relighting task, we provide both spatio-temporal conditioning and lighting control.

\paragraph{Spatio-Temporal Conditioning.}

To support spatio-temporal conditioning on the input frames, we follow previous works~\cite{relight_harmonization, instructp2p,atomovideo} and add additional input channels to the first convolution layer of the denoising U-Net.
This modification conditions the denoising process on the input video.

For temporal consistency over long sequences during testing, we adopt an iterative mechanism, detailed in Sec.~\ref{ssec:model_inference}.
As such, the model must distinguish the original input frames (\ie, albedo) from previous predictions (\ie, relit frames).
To achieve this, we introduce additional binary masks as input to the denoising U-Net.
These binary masks $M_{t}$ indicate the previously generated frames, \ie, they have all 1's for the frames from the previous window and 0's otherwise.
In summary, the input to the denoising U-Net is a concatenation of input latents, binary masks, and noise latents over time.

\paragraph{Lighting Control.}

To control the target lighting at test time, we condition the denoising U-Net on an HDR map.
Compared to other representations for lighting control, such as background images~\cite{relight_harmonization, iclight} or text prompts~\cite{iclight} recently used in conditional diffusion models, HDR maps capture a broader range of lighting scenarios and offer more precise control.

To provide the HDR map to the denoising U-Net, we first tokenize it.
Each directional ``light token" is computed by summing the intensities over a small local area in the HDR map.
We then embed these ``light tokens" into high dimensional ``light embeddings" using an multilayer perceptron (MLP), and concatenate them with positional encodings representing each light's average direction. The embeddings are then concatenated to form the light embedding $\mathcal{L}_e$.
$\mathcal{L}_e$ is transmitted to the denoising U-Net through cross-attention layers similar to text embeddings.
This conditioning design is key to achieving high-quality, controllable relighting results and enabling precise control as shown by our experiments. 

\subsubsection{Delighting Model}

The delighting diffusion model uses the same architecture as the relighting model.
It differs only in its input and conditioning signals.
Instead of taking a delit video and previous relit frames, it takes the original video and previous delit frames as input.
The delighting model does not use lighting control, since its target output is always a flat-lit albedo video.

\subsubsection{Long Sequence Prediction}
\label{ssec:model_inference}

Video diffusion models can only predict video frames with a fixed length (\eg $L$).
To support long-video inference, we adopt an iterative scheme following recent works~\cite{emo,atomovideo}.
We generate subsequent frames based on previous predictions.
Specifically, for our diffusion models, trained with frame length $L=30$, we replace the first $T=4$ frames in the input with previous predictions and update their masks, letting the model predict the subsequent $T-L=26$ frames.
During training, we randomly sample $T\in [0,4]$ and replace the input frames with ground truth.

\subsection{Learning Relighting on a Hybrid Dataset}
\label{ssec:model_training}

Training both delighting and relighting models would typically require a paired video dataset consisting of lit videos $V_l$ and their corresponding albedo videos $V_a$, as well as HDR map conditions $E_l$.
However, collecting such a video dataset is difficult and expensive in practice.

Instead, we use a more accessible hybrid video dataset created from static OLAT images and in-the-wild videos. We first introduce how we obtain training video datasets from both data sources and then present our training scheme, designed to best use the information from each.

\subsubsection{Hybrid Video Data Creation}
\vspace{-1mm}
\begin{figure}[t]
    \centering
    \includegraphics[width=\linewidth]{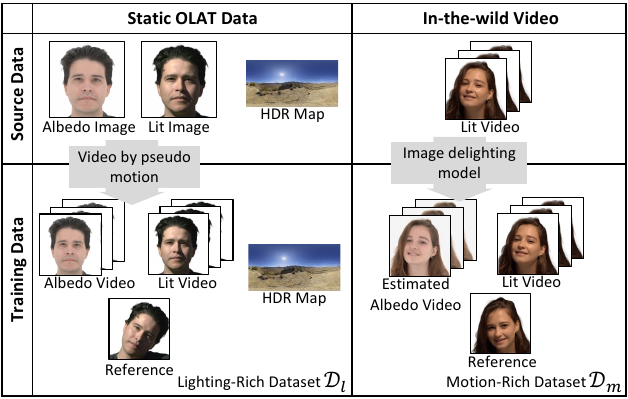}
    \caption{\textbf{Hybrid video dataset creation.}
    We add synthetic camera motion to static OLAT images, and apply the image delighting model to in-the-wild videos, creating our hybrid training data.}
    \label{fig:dataset}
    \vspace{-5mm}
\end{figure}

We build our hybrid dataset with two types of videos: videos with diverse illuminations but little motions (denoted as $\mathcal{D}_{l}$), synthesized from static OLAT images, and videos with diverse motions but unknown lighting (denoted as $\mathcal{D}_{m}$), obtained from in-the-wild videos.
A visualization of the raw source data and its corresponding processed training data is shown in Fig.~\ref{fig:dataset}.

\paragraph{Lighting-Rich Dataset $\mathcal{D}_{l}$.} $\mathcal{D}_{l}$ contains triples of paired video and HDR maps $\{V_l, V_a, E_l\}$.
To construct this dataset, we start from static OLAT images paired with matched flat-lit images.
We capture a group of subjects with diverse static expressions and poses from multiple views.
We use image-based lighting \cite{olat_paul,ibr_face} to obtain various relit versions of each expression/pose from their OLATs.
We generate the lit images using various HDR maps $E_l$.
For their corresponding albedo images, we directly use the flat-lit images \cite{total_relighting,lightpainter, holo_relight}.
Finally, to lift the data to the video domain we simulate camera motions (\eg, zooming, panning) on the images.
This lighting-rich dataset could be used to learn portrait relighting by itself.
However, the motions being limited to 2D camera movements, we demonstrate in our experiments that $\mathcal{D}_{l}$ is not enough to generalize to real in-the-wild motions.

\paragraph{Motion-Rich Dataset $\mathcal{D}_{m}$.} For $\mathcal{D}_{m}$, we use an in-the-wild video dataset~\cite{vfhq} containing 15,000 high-quality talking heads with diverse motion patterns.
As shown in Fig.~\ref{fig:dataset}, to create paired training samples, we first train an \emph{image} delighting model on $\mathcal{D}_{l}$ 
and then use this model, we predict pseudo albedos for each video, frame by frame. These albedo videos are thus not  temporarily consistent.
Details about the image delighting model can be found in the supplement.
This way, we have not only pseudo-ground truth for the delighting task but also pseudo-input with real ground truth for the relighting task, similar to~\cite{lightit}.

Thus, we obtain a dataset of in-the-wild video pairs $\mathcal{D}_{m} = \{V_l, V_a\}$, but without corresponding HDR map $E_l$.

\begin{figure}[t]
    \centering
    \includegraphics[width=\linewidth]{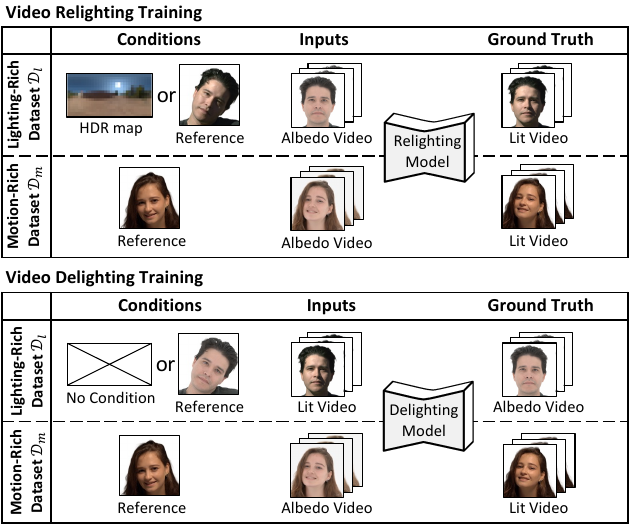}
    \caption{\textbf{Training with hybrid data.} To train both the relighting and delighting models, we use the hybrid dataset. We train the models on two tasks simultaneously: HDR-based condition (or no condition) on the OLAT data (\ie lighting-rich dataset $\mathcal{D}_{l}$) and reference-based appearance copy on both datasets (\ie motion-rich dataset $\mathcal{D}_{m}$ and lighting-rich dataset $\mathcal{D}_{l}$).}
    \label{fig:stages}
    \vspace{-5mm}
\end{figure}

\subsubsection{Training with Hybrid Data}
With a fully-labelled dataset of triples $\{V_l, V_a, E_l\}$, we could supervise both delighting and relighting models to learn the delighting mapping $\mathbf{D}: (V_l) \mapsto V_a$ and relighting mapping $\mathbf{R}_\text{hdr}: (V_a | \mathcal{L}_e) \mapsto V_l$, respectively.
Where $\mathcal{L}_e$ is the embedding of the HDR map $E_l$.

However, given our hybrid dataset, we face two issues.
First, we lack the HDR map $E_l$ to condition the relighting mapping $\mathbf{R}_\text{hdr}$ on the motion-rich dataset $\mathcal{D}_{m}$.
Second, for the delighting mapping $\mathbf{D}$, $\mathcal{D}_{m}$ only provides pseudo albedo, which is not temporally consistent enough to be used as supervision.
We thus propose to train both models on the auxiliary task of reference-based appearance copy.
This lets the networks learn from both datasets, combining accurate lighting control and improved temporal stability.

\paragraph{Reference-based Appearance Copy.}

To leverage the motion-rich dataset $\mathcal{D}_{m}$, we propose to train our network on an auxiliary task.
We choose the task of reference-based appearance copy~\cite{shu2018, shih2014}.
It presents similarities with relighting whilst not requiring known HDR maps.
Instead of using an HDR map to control lighting, for this task, we use a reference frame $F^i_l$ taken from a lit video $V_l$.
The network is then tasked to ``copy'' the appearance from the lit reference frame $F^i_l$ to the frames of the corresponding delit clip $V_a$. $F^i_l$ is randomly selected from $V_l$ and may not overlap with the training subsequences.
In practice, to condition the model on the reference frame $F^i_l$, we encode it as an embedding $\mathcal{L}_{ref}$ using a CNN encoder: $\mathcal{L}_{ref} = \mathcal{E}(F^i_l)$, thus leading to a reference-based appearance copy mapping $\mathbf{R}_\text{ref}: (V_a| \mathcal{L}_{ref}) \mapsto V_l$. To support both HDR- and reference-based lighting control, we adapt the model to accept different conditions using masks.
More details are in the supplement.
Next, we introduce how we train both video relighting and delighting using this auxiliary task.

\begin{figure*}[t!]
	\centering
	\includegraphics[ width=\textwidth]{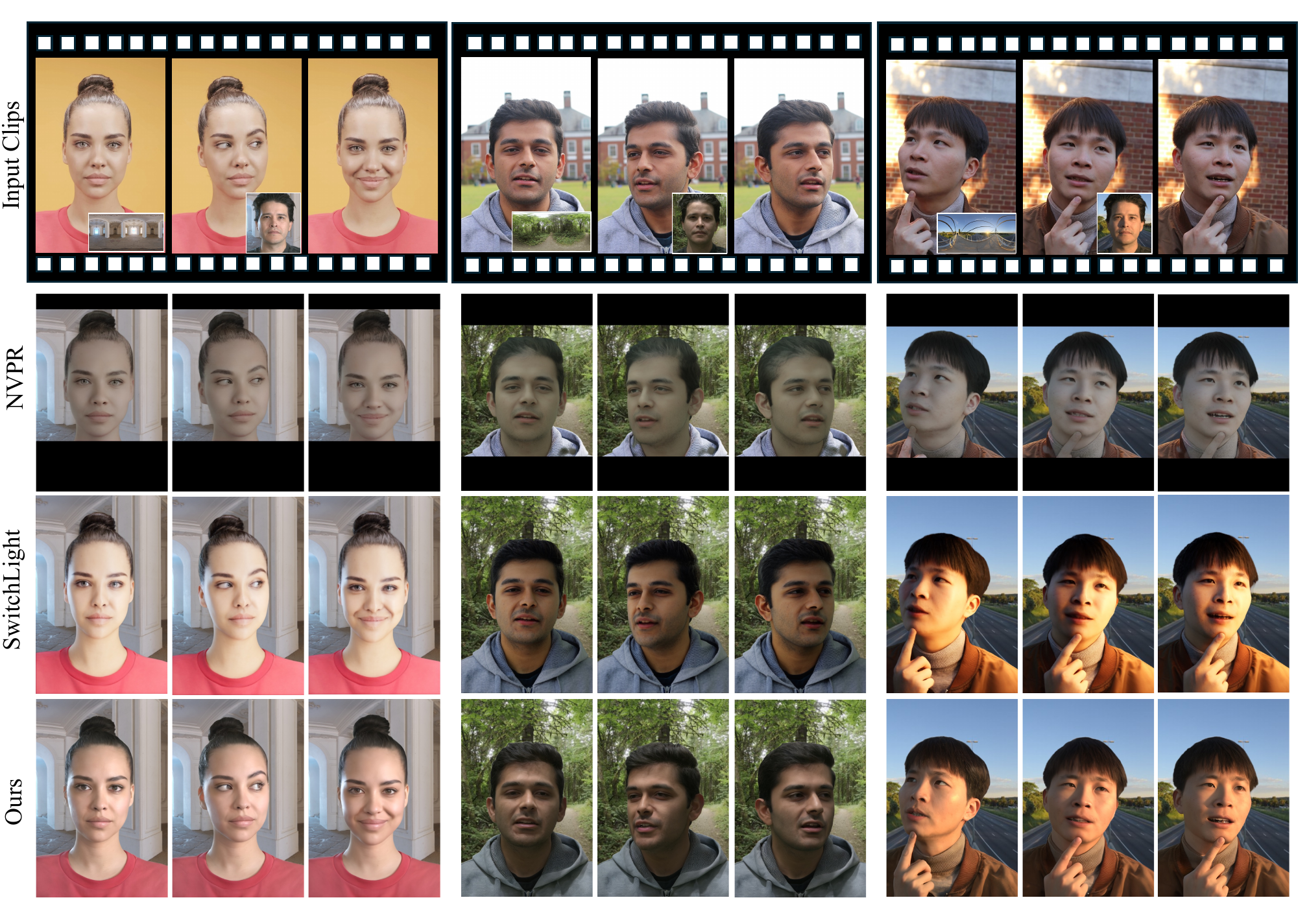}
\vskip-12pt	\caption{\textbf{Comparison against video relighting methods on in-the-wild portrait videos.}
For each sequence we show three input frames, with the target HDR map and reference image rendered with the same HDR map and OLAT data, both shown as inset.
Our method produces more faithful lighting effects and is robust to facial expression change (first column) and head motion (last two columns).}\label{fig:video-wild}
\vspace{-5mm}
\end{figure*}

\paragraph{Video Relighting.}
Our relighting model can now leverage both datasets by training HDR-based relighting and reference-based appearance copy simultaneously, as shown in Fig.~\ref{fig:stages} (Video Relighting).
Formally, $\mathbf{R} = \mathbf{R}_\text{hdr} \cup \mathbf{R}_\text{ref}: (V_a| \mathcal{L}_e \vee \mathcal{L}_{ref}) \mapsto V_l$.

For videos from the motion-rich dataset $\mathcal{D}_{m}$, we only use the reference-based conditioning, while we randomly condition on either or both for videos from the lighting-rich dataset $\mathcal{D}_{l}$. Training with the motion-rich dataset $\mathcal{D}_{m}$ through reference-based appearance copying effectively improves temporal consistency.

\paragraph{Video Delighting.}
For the relighting mapping, the motion dataset $\mathcal{D}_{m}$ provides temporally-inconsistent pseudo-albedo as input, but the relighting model is robust to it since the target is always a temporally consistent real video.
To train the delighting model $\mathbf{D}$, the mapping is inverted.
Training with pseudo-albedo from $\mathcal{D}_{m}$ as \emph{targets} leads to poor temporal consistency, which can affect the following relighting stage. We find that jointly training the delighting model to delight the lighting-rich dataset $\mathcal{D}_{l}$, while performing appearance copy on the motion-rich dataset $\mathcal{D}_{m}$, similar to the relighting model, yields satisfactory results, as the model can still learn temporal consistency through appearance copy. This process is shown as Video Delighting in Fig.~\ref{fig:stages}.
We hypothesize that the main benefit from the appearance copy task is to enforce temporal consistency.

%% file: sec/4_data.tex
\section{Data and Implementation Details}
\label{sec:data}

\paragraph{Data Preparation.} The hybrid dataset is generated from two sources: static OLAT captures and in-the-wild videos from VFHQ~\cite{vfhq}, which contains 15,000 high-quality talking head videos.
Our light stage is similar to~\cite{single_portrait_relighting,total_relighting} with 110 programmable LED lights and 36 frontal cameras.
We collect data from 67 participants, photographed with diverse head poses, facial expressions and accessories, resulting in a total of 71,280 OLAT sequences.
We reserve 7 subjects with varied appearances and genders for evaluation.
Following recent works~\cite{lightpainter, total_relighting, holo_relight}, we capture flat-lit images with all lights turned on, approximating diffuse albedo.
To obtain a high-quality dataset with diverse illuminations, we collect 769 HDR maps from PolyHeaven~\cite{PolyHeaven}, which are randomly paired with OLAT sequences to create the training dataset, using image-based relighting~\cite{olat_paul,ibr_face}.
600 HDR maps are randomly selected for training and the rest are used for testing. Our final static OLAT dataset contains about 4M rendered images.

\paragraph{Training Details.} We initialize both the relighting and delighting models with pre-trained SVD~\cite{stable_video_diffusion}.
A training batch consists of 8 video clips at resolution of 512$\times$768 pixels sampled from our hybrid dataset.
We train each model, with a two-stage training schedule.
In the first stage, we warm up the model by training on short sequences of 2 frames and optimize all model weights for 200K steps.
This speeds-up convergence and the model is quickly tuned to the relighting task.
In the second stage, we train with longer sequences of 30 frames over 50K iterations, only optimizing the temporal layers of our models.
We optimize both models towards v-prediction objectives~\cite{v-prediction} using the AdamW~\cite{adamW} optimizer with a learning rate of 1e-5. More data and training details can be found in the supplement.

%% file: sec/5_experiment.tex
\section{Experiments}
\label{sec:experiment}

We demonstrate the high-quality video relighting capability of \name through extensive evaluations. More video comparison results can be found in the supplement and on our~\href{https://www.eyelinestudios.com/research/luxpostfacto.html}{project page}. 

\paragraph{Evaluation Metrics.} We follow previous work~\cite{holo_relight,lightpainter} and use PSNR and SSIM~\cite{ssim} to evaluate fidelity and report LPIPS~\cite{lpips} and NIQE~\cite{NIQE} scores for perceptual quality.
All metrics are computed on the foreground subject, using pre-computed masks from~\cite{transparentbg}. 
\subsection{Comparison with State-of-the-Art Methods}

\subsubsection{Video Relighting}
We compare \name with the state-of-the-art video relighting methods NVPR~\cite{nvpr} and SwitchLight~\cite{switchlight} on in-the-wild portrait videos.
NVPR is a video relighting model trained on a large dynamic OLAT dataset.
SwitchLight~\cite{switchlight}, originally proposes an image relighting model.
Recently, the same authors deploy an improved video relighting model commercially, and we compare to this improved version.
Since we have no access to ground-truth relit videos, we present a qualitative comparison to both methods in Fig.~\ref{fig:video-wild}.
Our method produces, more realistic highlights and skin tones, finer details and better shadows.
In our supplement, we present more video comparisons, highlighting that our method is also more temporally stable.

\begin{figure}[t!]
	\centering
	\includegraphics[ width=\columnwidth]{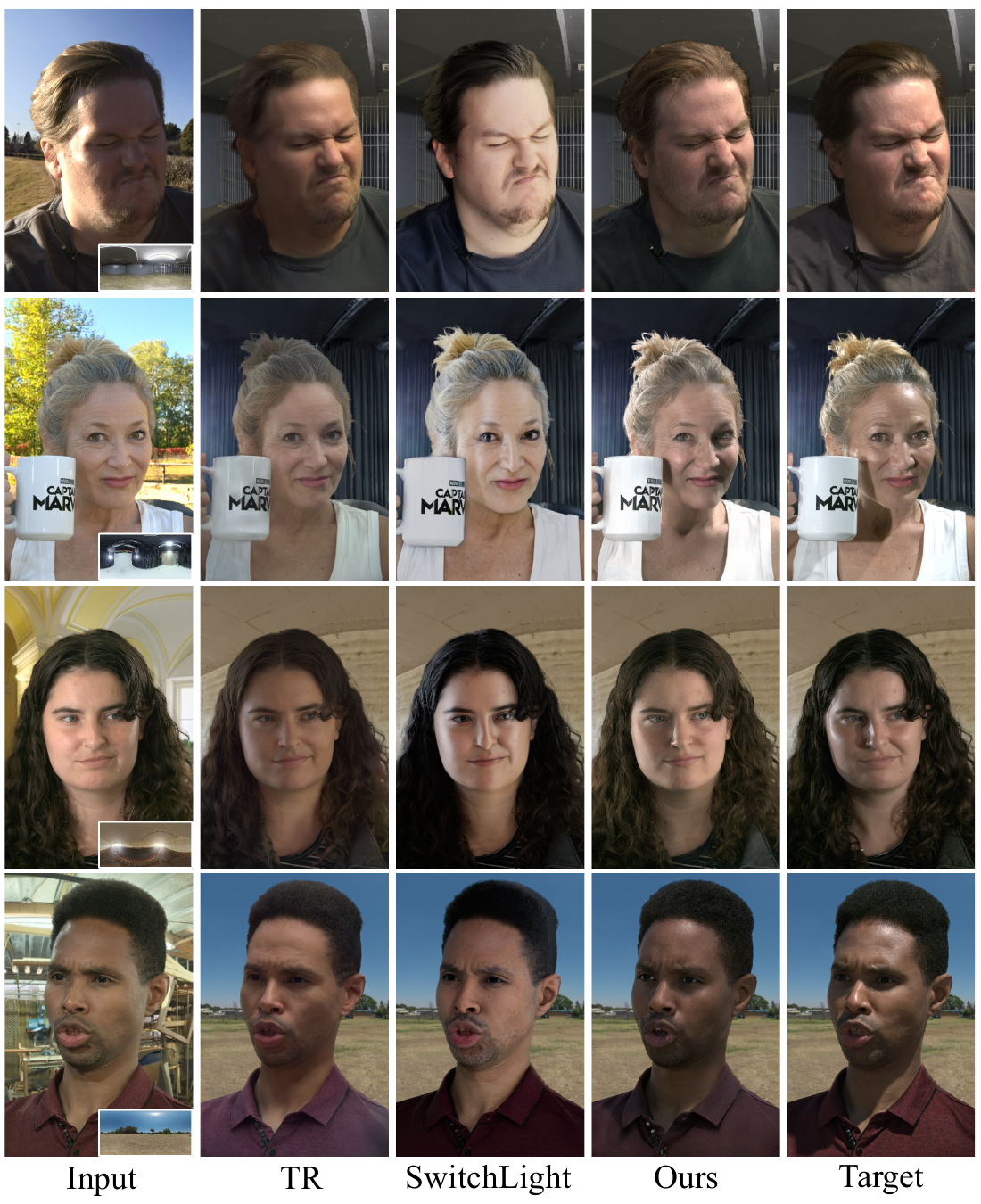}
\vskip-8pt	\caption{\textbf{Comparisons to image-based relighting methods on our test set.}
The first column shows the input and target HDR map. Subsequent columns show results from different methods and ground truth targets. Our results are more faithful and realistic.}\label{fig:fig_val_gt}
\end{figure}

\begin{figure}[t!]
	\centering
	\includegraphics[ width=\columnwidth]{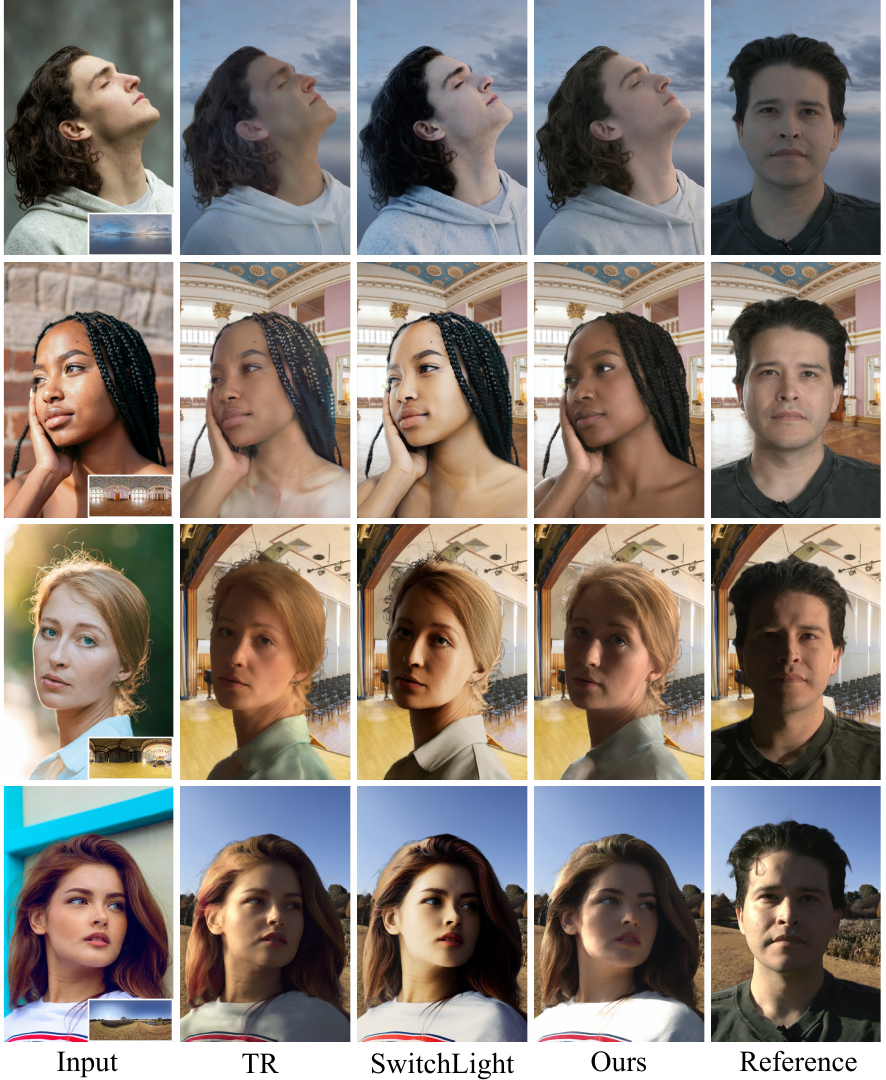}
\vskip-8pt	\caption{\textbf{Comparisons to image-based relighting methods on in-the-wild portrait images.} We present results similarly to Fig.~\ref{fig:fig_val_gt}, replacing the ground truth with an OLAT computed image as reference. Again, our method has better results with good tone preservation and details.}\label{fig:fig_val_ref}
\vspace{-2mm}
\end{figure}

\subsubsection{Image Relighting}
Besides video relighting, \name can also be used on \emph{single image} portraits, treating images as short static videos.
This allows to quantitatively evaluate our model's performance.
To do so, we compare \name with the state-of-the-art Total Relighting (TR)~\cite{total_relighting} method and the SwitchLight image model~\cite{switchlight}.
We conduct qualitative evaluations on in-the-wild images and both qualitative and quantitative evaluations on our test set, composed of left-out OLATs.
The quantitative results are presented in Tab.~\ref{tab:2d}.

\input{tables/image_eval}

While our method is trained with the added temporal consistency constraint, it outperforms both SoTA methods on all metrics by a fair margin.
In Fig.~\ref{fig:fig_val_gt} \&  ~\ref{fig:fig_val_ref}, we present qualitative results.
In Fig.~\ref{fig:fig_val_gt}, for which we use subjects from our test set, we observe our method produces results closer to the ground truth.
In Fig.~\ref{fig:fig_val_ref}, we use in-the-wild portraits and show one OLAT captured subject relit by the target lighting as reference.
 Our method respects skin tones better and produces more plausible highlights and shadows.

\begin{figure}[t!]
	\centering
	\includegraphics[ width=\columnwidth]{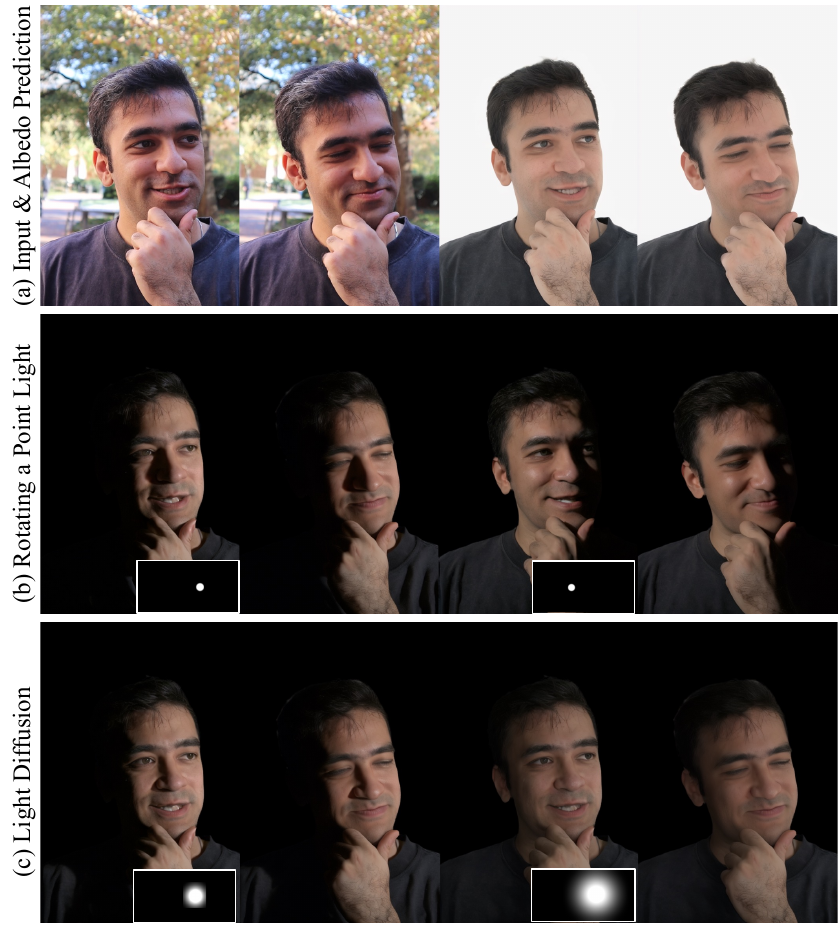}
\vskip-8pt	\caption{\textbf{Albedo and directionally lit relighting results.} Our method can predict (a) temporally consistent albedo, and (b) enables precise lighting control. Diffusing the light source effectively leads to softer shadows (c).}\label{fig:point_light}
\vspace{-2.5mm}
\end{figure}

\subsection{Additional Evaluations} We provide additional evaluations for \name to further demonstrate its capabilities.

\paragraph{Consistent Albedo Prediction.} Benefiting from our hybrid training strategy, \name predicts temporally consistent albedo channel as shown in Fig.~\ref{fig:point_light} (a).
The results show that our model removes the existing shading and preserves the facial details.  

\paragraph{Lighting Control.} \name enables precise lighting control thanks to our lighting injection mechanism.
We demonstrate its high controllability by relighting portraits using directional light sources.
As shown in Fig.~\ref{fig:point_light} (b), our results showcase complex light transport effects such as specular reflections and cast shadows coherent with the rotating light sources.
In Fig.~\ref{fig:point_light} (c), we make the directional light source more diffuse and show that our method renders realistic soft shadows. 

\begin{figure}[t!]
	\centering
	\includegraphics[ width=\columnwidth]{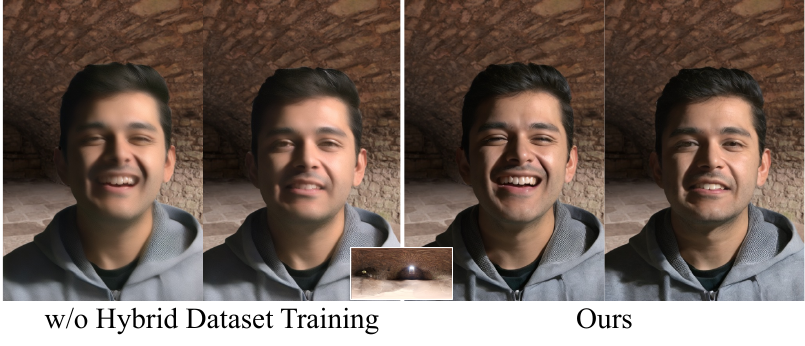}
\vskip-9pt	\caption{\textbf{Ablation study on hybrid dataset training.} Without hybrid training on the motion-rich in-the-wild videos, the model produces blurry results.}\label{fig:ab}
\vspace{-5.5mm}
\end{figure}

\input{tables/table_ab}

\subsection{Ablation Studies}
We conduct ablation studies on two core designs: lighting control module and hybrid dataset training strategy.
We first show that our lighting control mechanism is significantly better than using the commonly used CLIP image encoder~\cite{CLIP} for image conditioning (Tab.~\ref{tab:ab}).
Then, in Fig.~\ref{fig:ab} we show that with our hybrid dataset training strategy our model produces sharper and more photorealistic results compared to training only on the lighting-rich data $\mathcal{D}_l$.
Without using it (\ie training solely on OLAT dataset with simulated camera motions), the model cannot handle subject's movement well and tends to produce blurry results (Fig.~\ref{fig:ab}). More results can be found in the supplement.

%% file: tables/image_eval.tex
\begin{table}[t]
       
        \small
        \centering
        \caption{
        Quantitative evaluations against image-based relighting methods. Our method outperforms others in all evaluation metrics.
        }
         \vspace{-2mm}
        \tabcolsep=0.22cm
        \resizebox{\columnwidth}{!}{
                \begin{tabular}{l|cc|cc}
                        \hline
                        Methods      & LPIPS$\downarrow$   &NIQE$\downarrow$      & PSNR$\uparrow$  & SSIM$\uparrow$  \\ \hline \hline
                              TR~\cite{total_relighting}    &0.1794 &6.458 &22.44  &0.7793 \\ 
                              SwitchLight~\cite{switchlight} & 0.2129 & 7.166 & 19.87 & 0.7481 \\
                        
                        \textbf{Ours} & \textbf{0.1158}  & \textbf{5.653} & \textbf{24.62}  &\textbf{0.8278} \\ \hline
                        
        \end{tabular}}\label{tab:2d}
        \vspace{-5mm}
\end{table}

%% file: tables/table_ab.tex
\begin{table}[t]
       
        \small
        \centering
        \caption{
        Ablation study on lighting control module.
        }
         \vspace{-2mm}
        \tabcolsep=0.22cm
        \resizebox{\columnwidth}{!}{
                \begin{tabular}{l|cc|cc}
                        \hline
                        Methods      & LPIPS$\downarrow$   &NIQE$\downarrow$      & PSNR$\uparrow$  & SSIM$\uparrow$  \\ \hline \hline
                              Clip Image Enc. ~\cite{CLIP}    &0.1747 &5.831 &20.73  &0.7808 \\ 
                        
                        \textbf{Ours} & \textbf{0.1158}  & \textbf{5.653} & \textbf{24.62}  &\textbf{0.8278} \\ \hline
                        
        \end{tabular}}\label{tab:ab}
        \vspace{-4mm}
\end{table}

%% file: sec/6_conclusion.tex
\section{Conclusion}
\label{sec:conclusion}
We presented \name, a video relighting method that brings realistic and temporally consistent relighting to standard portrait videos in post-production.
By treating video relighting as an HDR map conditioned generation process, and fine-tuning a state-of-the-art video diffusion model, \name achieves generalized fine-grained lighting control of portrait videos.
\name only requires a set of static OLAT images and a larger pool of in-the-wild videos for training.
This hybrid dataset effectively provides relighting supervision for individual frames while fostering temporal stability across frames. Discussion of limitations and future work can be found in the supplement.

\noindent\textbf{Acknowledgements} We extend our gratitude to Jeffrey Shapiro for his initiative and ongoing support; Jennifer Lao for organizing the capture procedures; Mike Farris, John Milward, Kevin Izumi, and JT Parino for stage operations; Connie Siu, Kevin Williams, Amir Shachar, and Alex Chun for stage coordination and data organization; Sunny Koya for technical support; the software department for maintaining the stage software; and Brian Tong for helping prepare the demonstration video.

%% file: supp.tex




\maketitlesupplementary

\begin{figure}[t!]
	\centering
	\includegraphics[ width=\columnwidth]{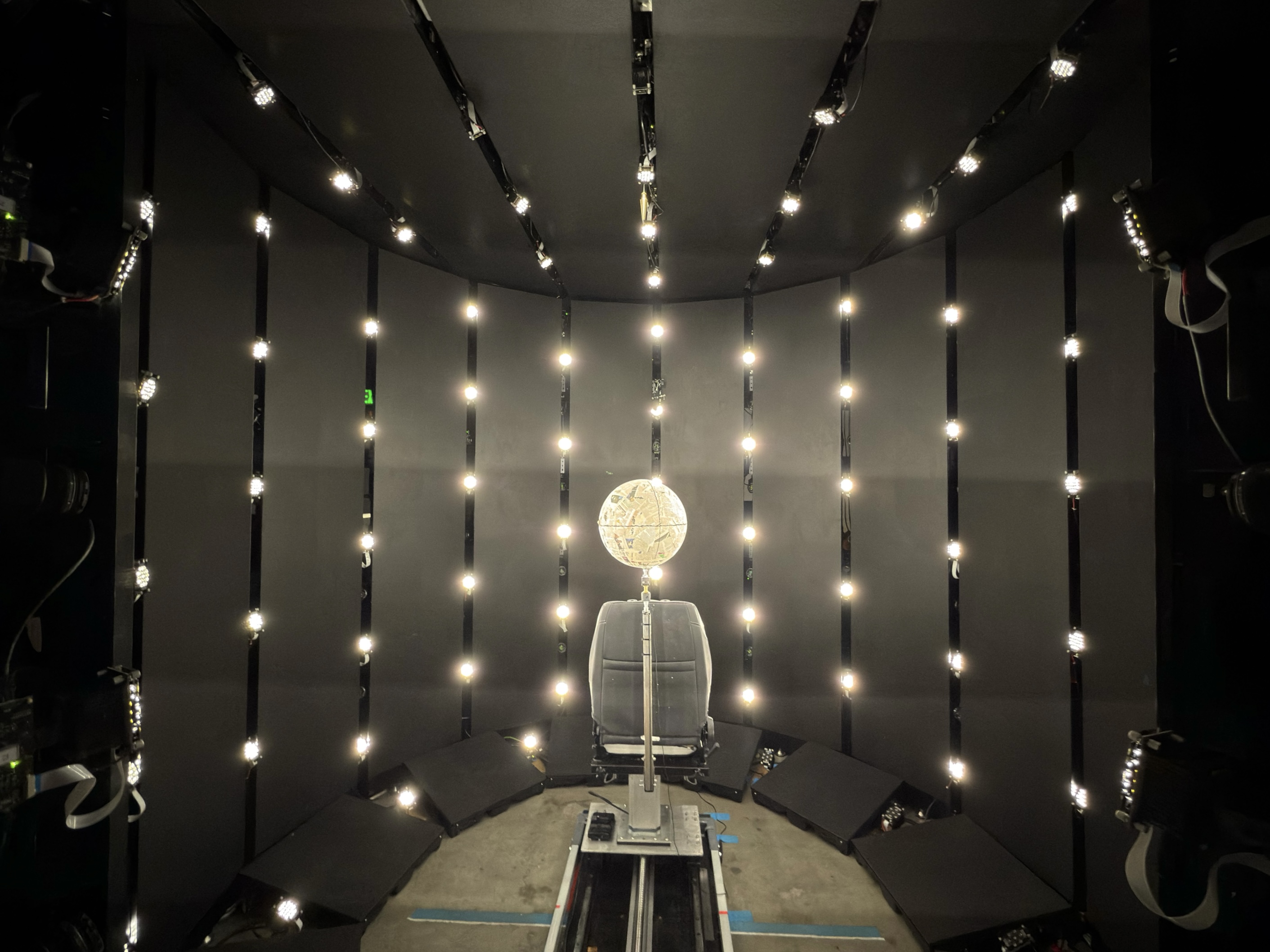}
\vskip-8pt	\caption{An illustration of our light stage.}\label{fig_supp:vps05}
\vspace{-1mm}
\end{figure}
\begin{figure}[t!]
	\centering
	\includegraphics[ width=\columnwidth]{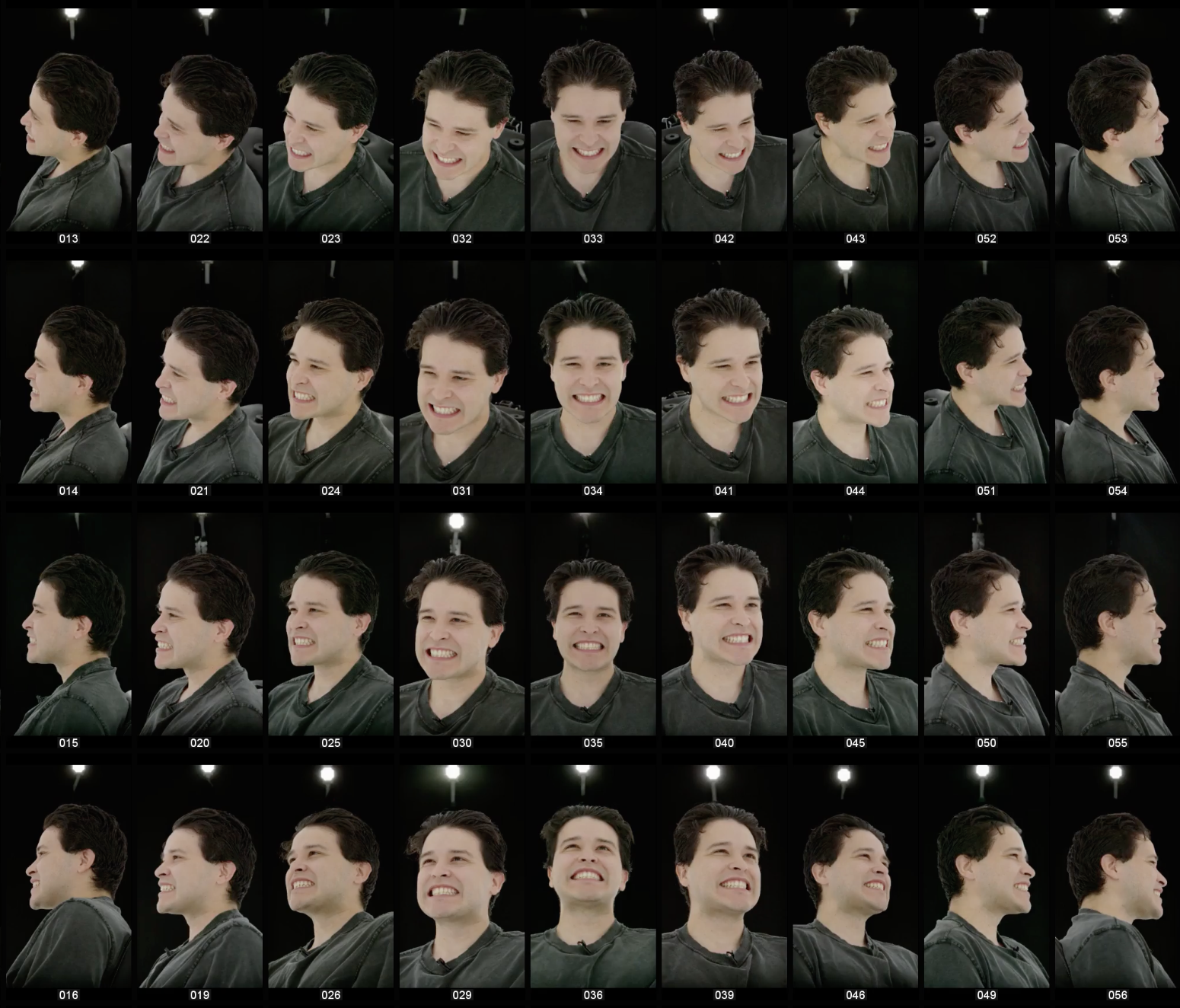}
\vskip-8pt	\caption{Examples of captured views using 36 frontal cameras.}\label{fig_supp:frontal_views}
\vspace{-1mm}
\end{figure}

\begin{figure}[t!]
	\centering
	\includegraphics[ width=\columnwidth]{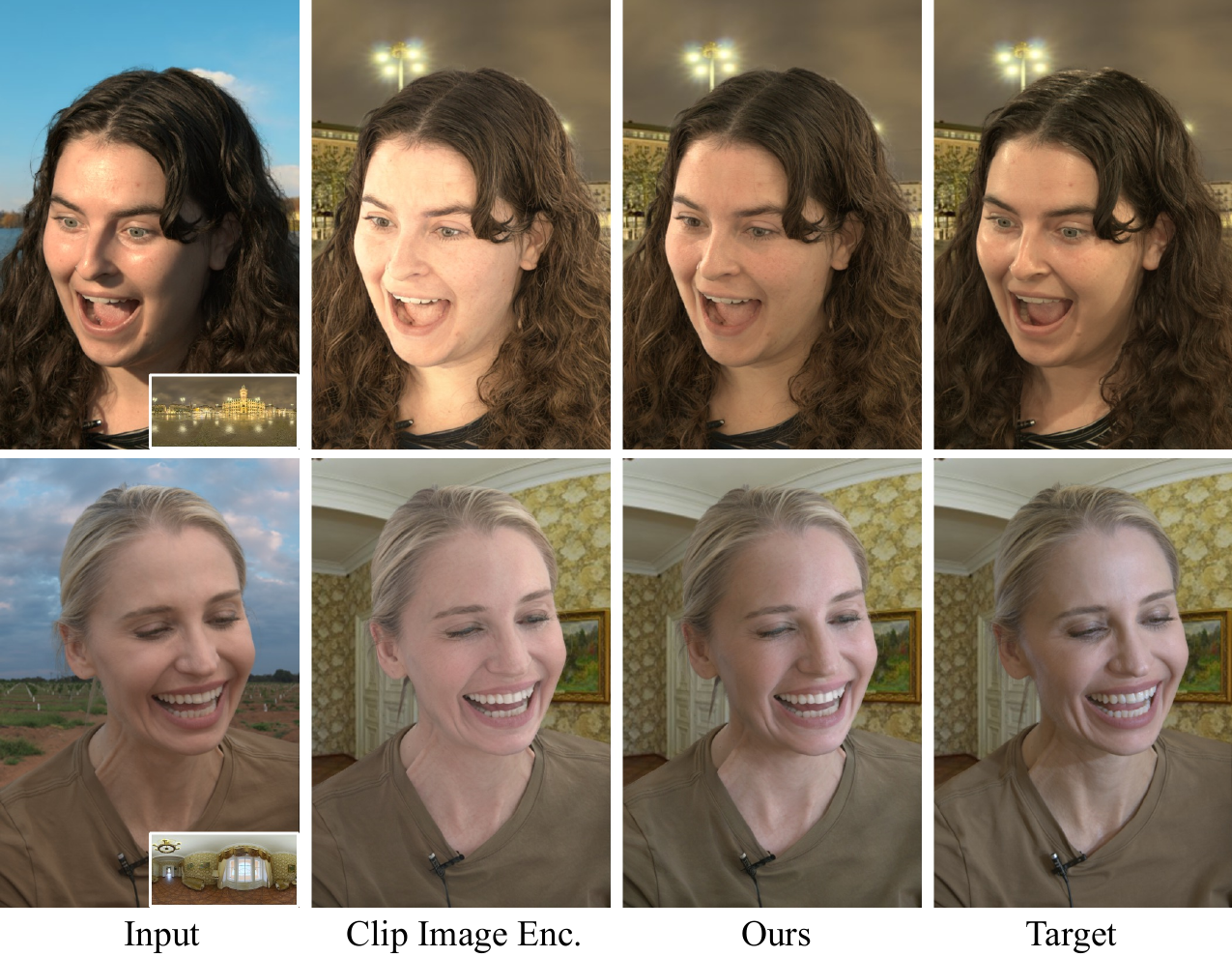}
\vskip-8pt	\caption{Visual comparisons for the ablation study on lighting control module. Compared to our design, the common used CLIP-based image encoding~\cite{CLIP,ye2023ip,zhao2024uni} cannot accurately capture the lighting intensity and directions in an HDR map and thus fails to enable precise lighting control. In contrast, our approach can produce high-quality lighting effects that follow the given HDR map.}\label{fig_supp: ablation}
\vspace{-1mm}
\end{figure}

\section{Video Demonstration}
We encourage readers to view the provided~\href{https://www.eyelinestudios.com/research/luxpostfacto.html}{supplemental
video}, which contains video results and comparisons, for a more comprehensive illustration of
the relighting quality of \name.

\begin{figure*}[t!]
	\centering
	\includegraphics[ width=\textwidth]{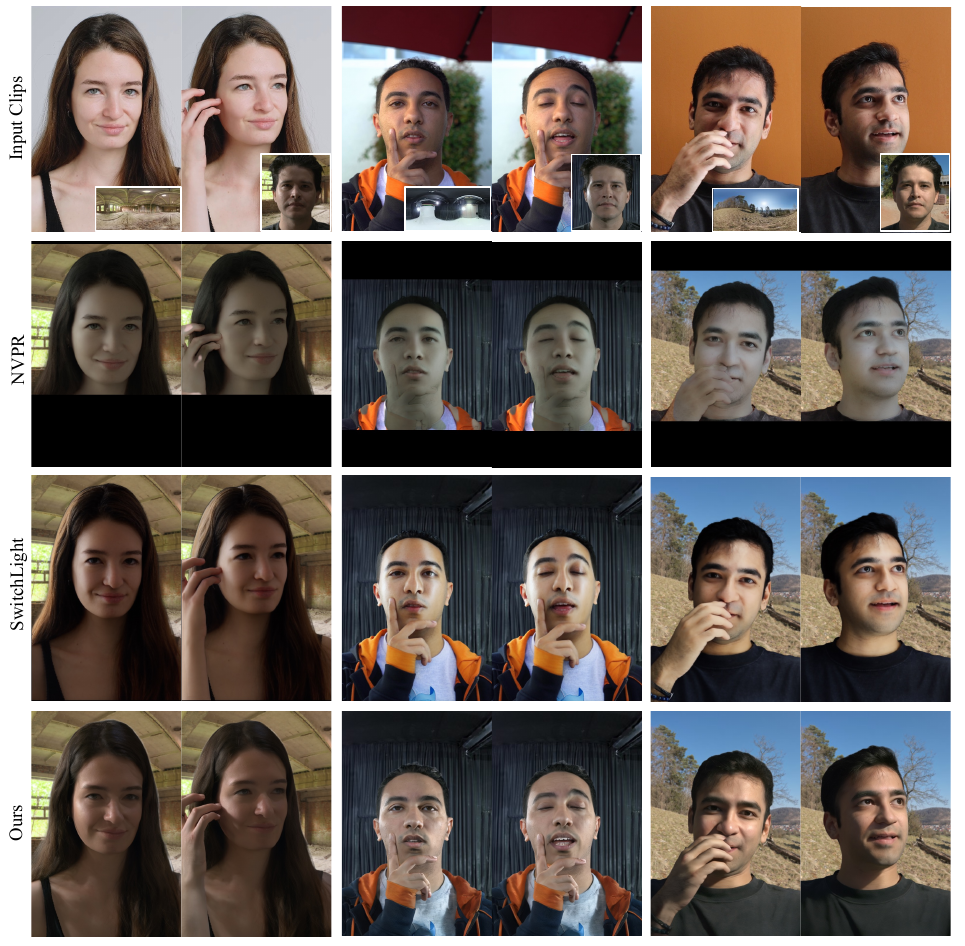}
\vskip-8pt	\caption{Visual comparisons with video relighting methods on in-the-wild portrait videos. We compare our method with NVPR~\cite{nvpr} and SwitchLight~\cite{switchlight}.}\label{fig_supp:video}
\vspace{-3mm}
\end{figure*}

\begin{figure*}[t!]
	\centering
	\includegraphics[ width=\textwidth]{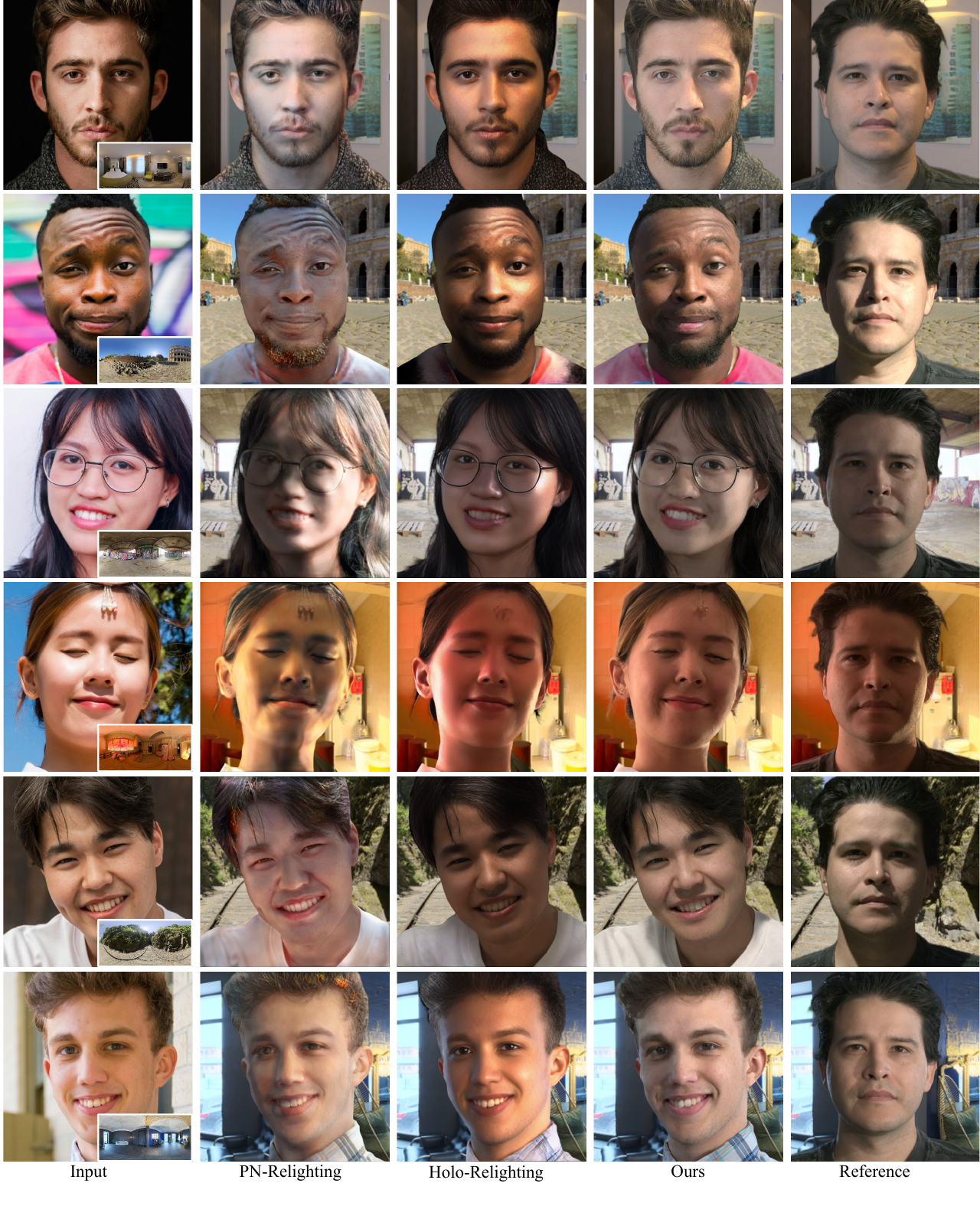}
\vskip-10pt	\caption{Visual comparisons on in-the-wild image relighting. We compare our method with PN-Relighting~\cite{smallot} and Holo-Relighting~\cite{holo_relight}. Both approaches are designed for
$512 \times 512$ face crops. Therefore, we report results on this region-of-interest for all methods for a fair comparison.}\label{fig_supp:face_crop}
\vspace{-2mm}
\end{figure*}

\begin{figure*}[t!]
	\centering
	\includegraphics[ width=\textwidth]{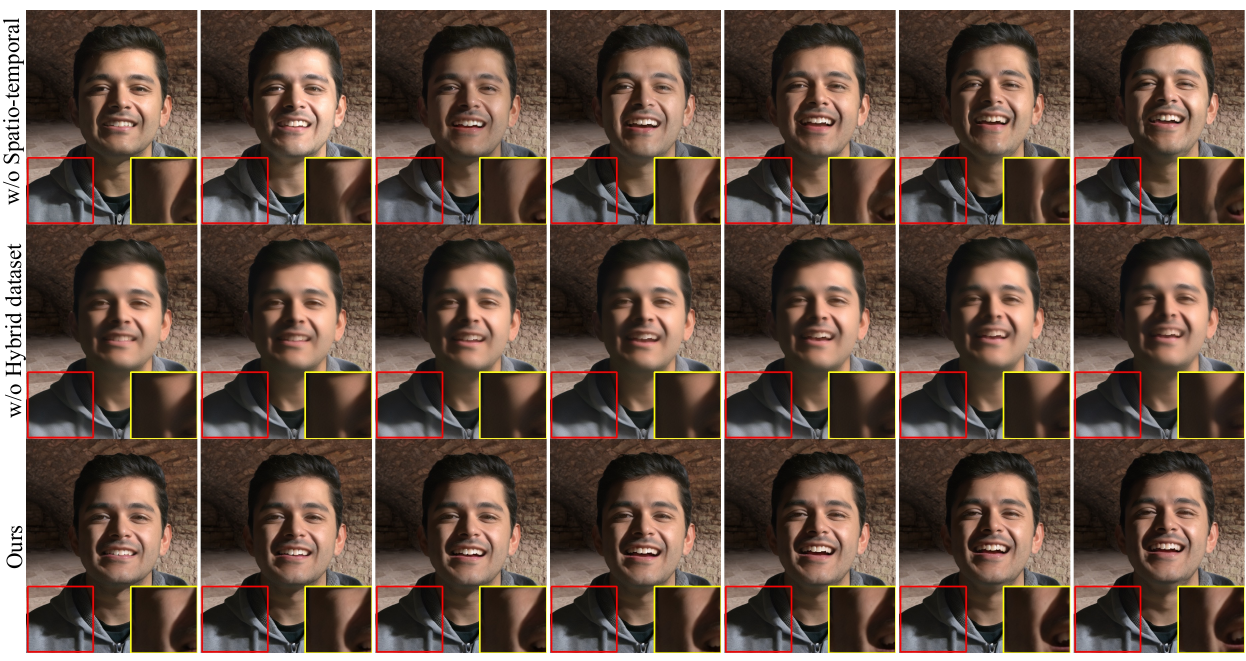}
\vskip-10pt	\caption{Additional visual evaluation on temporal consistency.}\label{fig_supp:temp}
\vspace{-1mm}
\end{figure*}

\begin{figure*}[t!]
	\centering
	\includegraphics[ width=\textwidth]{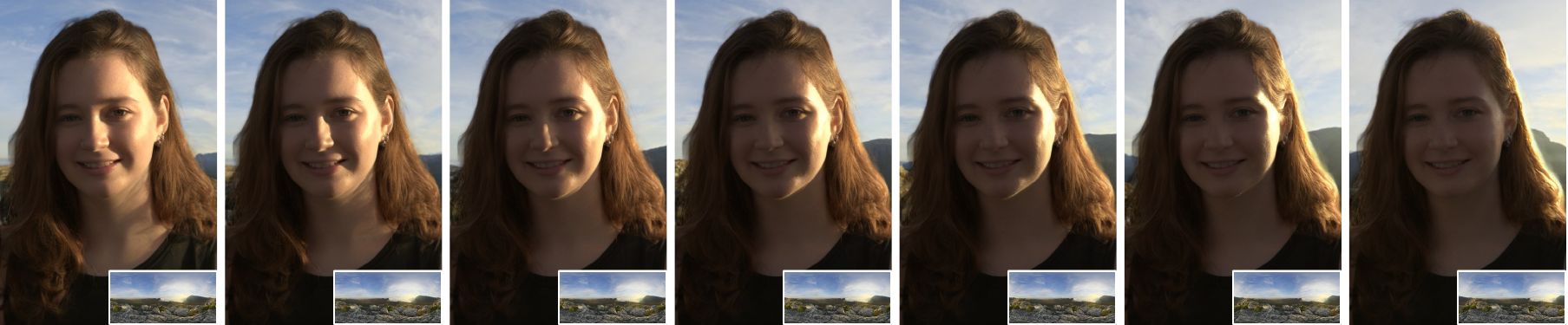}
\vskip-8pt	\caption{Relighting results under a rotating HDR map.}\label{fig_supp:rotate}
\vspace{-1mm}
\end{figure*}

\begin{figure}[t!]
	\centering
	\includegraphics[ width=\columnwidth]{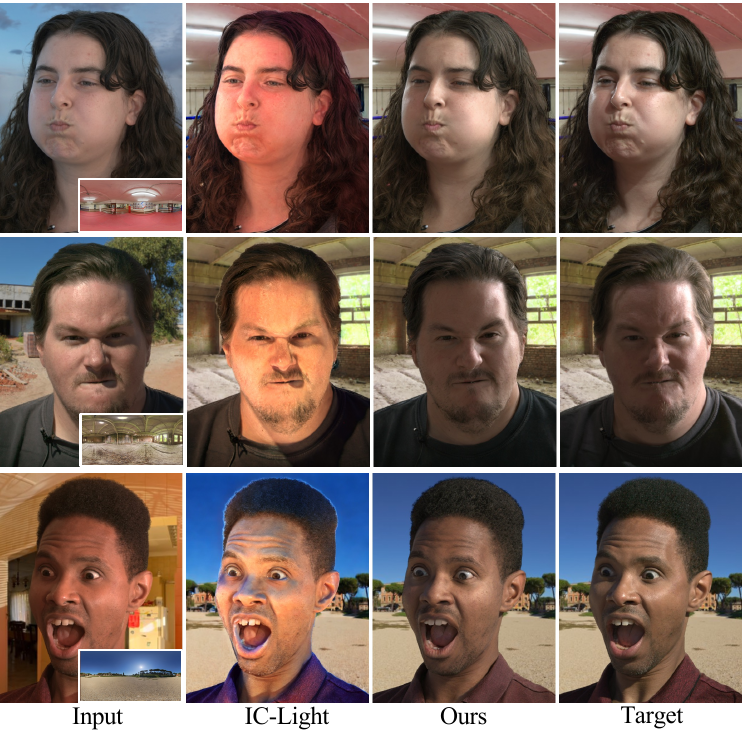}
\vskip-10pt	\caption{Visual comparisons to background-based relighting method IC-Light~\cite{iclight}. IC-Light produces results with artifacts and struggles with synthesizing precise lighting effects.}\label{fig_supp:ic_comp}
\vspace{-1mm}
\end{figure}

\begin{figure}[t!]
	\centering
	\includegraphics[ width=\columnwidth]{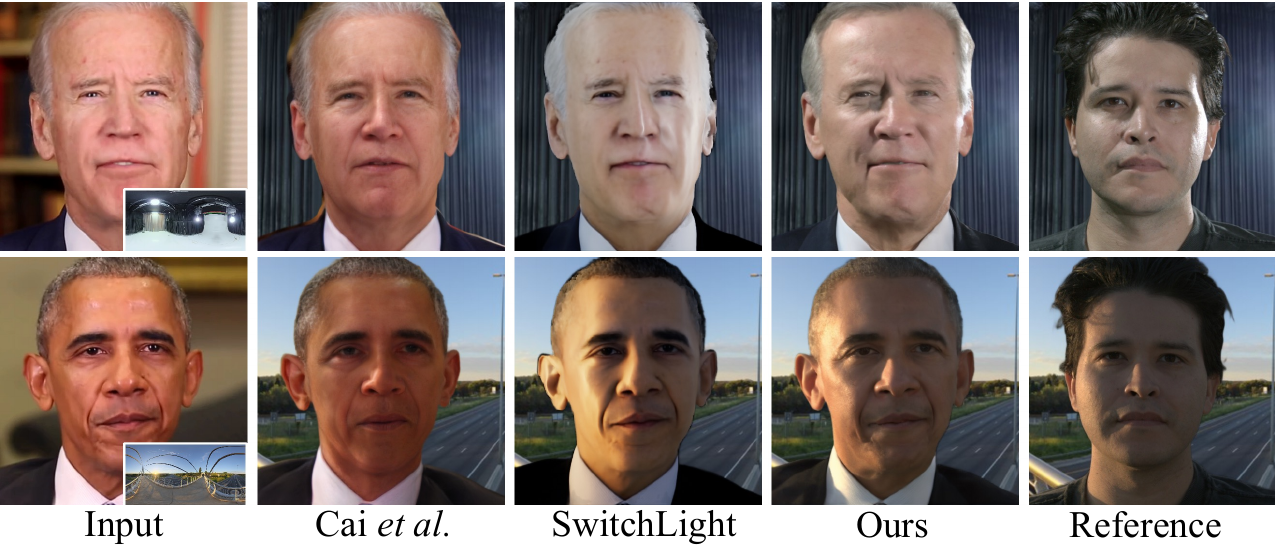}
\vskip-10pt	\caption{Visual comparisons to Cai~\etal~\cite{cai_etal} and SwitchLight~\cite{switchlight} on the INSTA dataset~\cite{insta}.}\label{fig_supp:insta}
\vspace{-2mm}
\end{figure}

\section{Additional Implementation Details}
\paragraph{Reference Frame Encoding.} \name is jointly trained with HDR-based relighting and reference-based appearance copy. To condition the model on a reference frame, we adopt a simple CNN encoder that encodes the image (512 $\times$ 512 $\times$ 3) into a feature map of 32 $\times$ 32 $\times$ 768. We reshape it as a list of image embeddings (\ie 1024 $\times$ 768) and append them after the lighting embeddings. When one conditioning (\eg HDR-based) is used, we deactivate the other conditioning (\eg reference-based) by replacing their embeddings with ``null" embeddings. 

\paragraph{Image Delighting Model.} We use an image delighting model to create paired training samples for the motion-rich dataset $\mathcal{D}_{m}$. We implement this model based on \textit{Stable Diffusion (SD)}~\cite{latent_diff}. The model is extended to be spatially conditioned on an input image by adding additional input channels to the first convolution layer of the denoising U-Net, similar to our video model, and the text embeddings are replaced by ``null" embeddings. We initialize the model weight from the pre-trained SD 1.5~\cite{SD15}, and supervisely train the model on our static OLAT dataset. We optimize the model towards v-prediction objectives~\cite{v-prediction} with a learning rate of 1e-5. The training stops after 200K steps.

\paragraph{More Training Details.} To support autoregressive inference for long sequence, we randomly sample $T\in [0,4]$ and replace the first $T$ input frames with ground truth during training. This allows the model to learn to generate subsequent frames based on previous predictions, therefore enhances temporal consistency across prediction windows. In our implementation, we sample $T=0$ with $p=0.5$ and other values equally with $p=0.125$. Our method is implemented using PyTorch and trained on 8 NVIDIA A100 GPUs. During testing, results are generated using DDIM~\cite{DDIM} sampler with 30 diffusion steps.

\paragraph{Light Stage and Rendering Details.}
We capture our static OLAT data using a light stage~\cite{olat_paul}. Specifically, the stage is configured as a cylindrical rig, equipped with 110 programmable LED lights and 75 Z-CAM e2 cinema cameras. We provide an illustration of the stage in Fig.~\ref{fig_supp:vps05}. We use 36 frontal cameras for this project. Examples of the captured views are provided in Fig.~\ref{fig_supp:frontal_views}. The stage has a diameter 2.7m and is 2.5m tall. The OLAT images are captured at 4K resolution. We cropped the upper body region and resize it to a resolution of $512\times 768$ for training. During rendering, we randomly pair each OLAT sequence with multiple HDR maps and obtain lit images using image-based relighting~\cite{olat_paul, ibr_face}. To diversify the illuminations, we augment an HDR map by randomly rotating it. Following~\cite{holo_relight}, we further add the original OLAT images into our rendered dataset. 

The use and collection of the OLAT data were reviewed and approved by the Institutional Review Board (IRB) and informed consent was obtained from all participants. 

\input{tables/table_temp}

\section{More Results for Ablation Study}
We provide visual results for the ablation study on lighting control in Fig.~\ref{fig_supp: ablation}. As shown, commonly used CLIP-based image encoding~\cite{CLIP, ye2023ip, zhao2024uni} cannot enable precise lighting control, whereas our lighting conditioning approach can produce high-fidelity lighting effects that follow the given HDR map.  

 We also provide additional evaluation on temporal consistency. We conduct ablation study on two key designs: (1) the conditional video diffusion model (spatio-temporal design), which is trained using (2) hybrid dataset training strategy. These two designs together enable temporally consistent and high-quality relighting. We report results in Fig.~\ref{fig_supp:temp} using a sequence of frames, and Tab.~\ref{tab:ab_temp} using the image quality metric NIQE~\cite{NIQE} and temporal metrics lighting error (LE), light instability (LI), LPIPS~\cite{lpips} between two adjacent frames and warping error (WE). Without spatio-temporal (\ie video) design, the corresponding image diffusion model produces flickering lighting effects (see shadows on shoulder and cheek). With video modeling but without hybrid dataset training, the resulting video model (solely trained on OLAT simulated data $\mathcal{D}_{l}$) produces temporally smooth but blurry results. 

\section{Relighting under Rotating HDR Maps} To further demonstrate the effectiveness of the lighting control module, we report relighting results under a rotating HDR map. 
As shown in Fig.~\ref{fig_supp:rotate}, our method can faithfully render lighting effects following the rotated HDR maps.

\section{More Comparison Results}
In Fig.~\ref{fig_supp:video}, we provide more visual comparisons against video relighting methods~\cite{nvpr,switchlight} on in-the-wild portrait videos. For NVPR~\cite{nvpr}, we acquire the results from the authors as their code is not available. For SwitchLight~\cite{switchlight}, we obtain their results by using their commercial application~\cite{switchlight_desk}.

In Fig.~\ref{fig_supp:face_crop}, we provide additional comparisons with two state-of-the-art face relighting method PN-Relighting~\cite{smallot} and Holo-Relighting~\cite{holo_relight}. PN-Relighting also uses the concept of data mixing but for a different goal (i.e. improving image relighting quality and albedo prediction) and via a different self-supervision approach. In contrast, we use data mixing for learning temporal consistent video relighting. The results for Holo-Relighting~\cite{holo_relight} are acquired from their authors as the source code is not available. Both approaches are designed for $512 \times 512$ face crops. Therefore, we report results on this region-of-interest for all methods for a fair comparison. Our methods generate more faithful relighting results, and the produced lighting effects are more consistent to the lighting effects in reference images. 

In Fig.~\ref{fig_supp:ic_comp}, we provide additional comparisons with background-based relighting method IC-Light~\cite{iclight} for image relighting. Compared to our method, IC-Light produces artifacts and fails to render precise lighting effects specified in the target HDR map.
\input{tables/table_ic}
\input{tables/table_insta}
In Tab.~\ref{tab:ic}, we report quantitative comparison with PN-Relighting~\cite{smallot} and IC-Light~\cite{iclight} on our test set. 

In Fig.~\ref{fig_supp:insta} and Tab.~\ref{tab:insta}, we additionally compare our method with Cai~\etal~\cite{cai_etal} and SwitchLight~\cite{switchlight} on the INSTA dataset~\cite{insta}. Note that INSTA dataset is designed for avatar reconstruction rather than evaluating video relighting performance. It may not best reflect the relighting capability as 1. it only contains a small number of subjects with limited input lighting and lack of large motions; 2. the videos are compressed, resulting in smoothed facial details in the input frames. On this dataset, our method achieves the best results both in terms of relighting quality and temporal consistency. 

\section{Limitations and Future Work}
\name is not without limitations. First, although our model can robustly handle most accessories, we found a few challenging cases where accessories, such as the decorative hairpiece shown in Fig.~\ref{fig_supp:face_crop}
 (row 4, column 4), partially occlude the face. In such scenario, the model may not perfectly preserve the accessory's details. This is because our training dataset lacks examples of faces with such occlusions, making it difficult for the model to handle this unseen case effectively. Second, as our model learns to synthesize lighting from the OLAT renderings, it can only generate lighting effects that can be represented by the light stage. Similar to previous methods~\cite{total_relighting, holo_relight, lightpainter, single_portrait_relighting, switchlight}, some challenging lighting effects (\eg foreign shadows) cannot be produced by our approach. Third, \name relies on video diffusion models to generate relit videos. The iterative nature of the diffusion process makes it challenging to apply our method for real-time applications. Further improving run-time efficiency might be a very interesting direction for future work. Some possible solutions include designing more efficient architectures~\cite{mobilediffusion} or exploring distillation techniques~\cite{consistencymodels,onestepdiff} to reduce sampling steps. We leave this direction for future work. Finally, due to GPU memory constraint, we train our model at a resolution of $512\times 768$. To support higher-resolution generation, one possible way is to utilize an off-the-shelf super-resolution model (\eg ~\cite{videosuperres, upscalevideo}) as a post-processing step. We leave such exploration for future work. 

\section{Potential Negative Social Impacts}
This method is designed to facilitate content creators to create creative and compelling lighting
in portrait videos. However, we acknowledge its potential misuse, such as creating deepfakes or misleading videos. Our work is developed to support positive and creative applications. To mitigate misuse of our relighting method, we advocate for responsible usage, clear content labeling and implementing robust detection mechanisms.


%% file: tables/table_temp.tex
\begin{table}[t]
       
        \small
        \centering
        \caption{
        Quantitative evaluation on temporal consistency.
        }
         \vspace{-2mm}
        \tabcolsep=0.1cm
        \resizebox{\columnwidth}{!}{
                \begin{tabular}{l|c|cccc}
                        \hline
                        Methods      & NIQE$\downarrow$   &LE$\downarrow$      & LI$\downarrow$  & LPIPS (temp.)$\downarrow$  & WE$\downarrow$ \\ \hline
                        \hline
                        w/o Spatio-temporal & {5.471} & 0.5542 &0.0796 &0.0450 & 0.0013 \\
                        w/o Hybrid dataset. &6.639 & 0.5233 &0.0387 &0.0081 &0.0001 \\
                        \hline
                        \textbf{Ours} & \textbf{5.462} & \textbf{0.4978} & \textbf{0.0350}  &\textbf{0.0073} &\textbf{0.0001}\\
                        \hline
                    
        \end{tabular}}\label{tab:ab_temp}
        \vspace{-2mm}
\end{table}

%% file: tables/table_ic.tex
\begin{table}[t]
        \small
        \centering
        \caption{
        Quantitative comparison with PN-Relighting and IC-Light on our test set.
        }
         \vspace{-1mm}
        \tabcolsep=0.3cm
        \resizebox{\columnwidth}{!}{
                \begin{tabular}{l|cc|cc}
                        \hline
                        Methods      & LPIPS$\downarrow$   &NIQE$\downarrow$      & PSNR$\uparrow$  & SSIM$\uparrow$  \\ \hline \hline
                              PN-Relighting~\cite{smallot}    &0.2486 &7.799 &17.15  &0.7373 \\ 
                              IC-Light~\cite{iclight} &0.2519 & 6.996 &16.12 &0.7315 \\
                        \hline
                        \textbf{Ours} & \textbf{0.1158}  & \textbf{5.653} & \textbf{24.62}  &\textbf{0.8278} \\ \hline
                        
        \end{tabular}}\label{tab:ic}
        \vspace{-2mm}
\end{table}

%% file: tables/table_insta.tex
\begin{table}[t]
       
        \small
        \centering
        \caption{
        Quantitative evaluation on INSTA dataset.
        }
         \vspace{-3mm}
        \tabcolsep=0.2cm
        \resizebox{\columnwidth}{!}{
                \begin{tabular}{l|c|cccc}
                        \hline
                        Methods      & NIQE$\downarrow$   &LE$\downarrow$      & LI$\downarrow$   & LPIPS (temporal)$\downarrow$  & WE$\downarrow$ \\ \hline \hline
                        Cai \etal & 6.582 & 0.6521 & 0.1495  &0.0206 &0.0002 \\
                        SwitchLight & 7.107 & 0.5987 & 0.0822  & 0.0128  & 0.0001\\
                        \hline
                        \textbf{Ours} & \textbf{5.953} & \textbf{0.5239} & \textbf{0.0451}  & \textbf{0.0092} & \textbf{0.0001} \\
                        \hline
        \end{tabular}}\label{tab:insta}
        \vspace{-4mm}
\end{table}